\documentclass{jetpl}
\twocolumn

\usepackage[dvipsnames]{xcolor}
\usepackage{tabularx}
\newcolumntype{Y}{>{\centering\arraybackslash}X}
\newcolumntype{C}[1]{>{\centering\arraybackslash}p{#1}}
\usepackage{pgfplots}
\pgfplotsset{compat=1.18}

\usepackage{caption}
\usepackage{subcaption}
\usepackage{multirow}
\usepackage{hyperref}
\usepackage[numbers,sort&compress]{natbib} 

\usepackage[normalem]{ulem}

\lat

\title{Ultra-high energy event KM3-230213A as a cosmogenic neutrino in light of minimal UHECR flux models}

\rtitle{Ultra-high energy event KM3-230213A as a cosmogenic neutrino in light of minimal UHECR flux models}

\sodtitle{Ultra-high energy event KM3-230213A as a cosmogenic neutrino in light of minimal UHECR flux models}

\author{M.\,Yu.\,Kuznetsov$^{\,a}$,
N.\,A.\,Petrov$^{\,a,\,b}$,
Y.\,S.\,Savchenko$^{\,b,\,c}$
}

\rauthor{M.\,Yu.\,Kuznetsov,
N.\,A.\,Petrov,
Y.\,S.\,Savchenko
}

\sodauthor{Kuznetsov, Petrov, Savchenko}

\address{~\\$^a$Institute for Nuclear Research of the Russian Academy of Sciences, 60th October Anniversary Prospect\,7a, Moscow, 117312, Russia\\~\\
$^b$Budker Institute of Nuclear Physics of Siberian Branch Russian Academy of Sciences, Akademika Lavrentieva prospect\,11, Novosibirsk, 630090, Russia\\~\\
$^c$Department of Physics, Novosibirsk State University, Pirogova street\,1, Novosibirsk, 630090, Russia}

\abstract{Recently, the KM3NeT experiment reported the detection of a neutrino with exceptionally high energy $E = 220$~PeV, whose origin remains unclear. 
The corresponding value of the neutrino flux is in tension with the results of other high-energy neutrino experiments. 
In this study, we discuss the possibility that this neutrino is cosmogenic, i.e., produced by ultra-high energy cosmic rays (UHECR) during their propagation through the intergalactic medium. 
We adopt the UHECR flux models derived by the Telescope Array experiment, which features a predominantly light mass composition. 
We show that the predictions of the cosmogenic neutrino flux in these models are consistent with the measurements of the KM3NeT-only and with that of the ``global neutrino observatory'' at approximately $2\sigma$ level. 
Notably, this result is achieved in a minimal version of the UHECR flux models, that assume one source population with a standard cosmological evolution.
We also estimate the corresponding cosmogenic gamma-ray flux and show that it is consistent with Fermi-LAT Isotropic Diffuse Gamma-Ray Background (IGRB) measurements and ultra-high energy gamma-ray limits; the improvement of the latter can probe these predictions in future.}

\begin{document}

\maketitle

\paragraph{1. Introduction.}\label{sec:intro}
On 13 February 2023, the KM3NeT experiment observed an exceptionally high-energy event called KM3-230213A~\cite{KM3NeT:2025npi}. This event is compatible with a muon with an estimated energy of 120~PeV, which, in turn, was produced by an even more energetic neutrino that interacted in the vicinity of the detector. Namely, the estimate for the neutrino energy is 220~PeV and it lies within the $72\; \mbox{PeV} - 2.6\; \mbox{EeV}$ energy range at the 90\% confidence level. This makes the KM3-230213A neutrino the most energetic neutrino ever detected.
It is interesting that other high-energy neutrino experiments with much larger exposures have not observed any neutrinos at these energies. Namely, the upper limits on the ultra-high energy~(UHE) neutrino flux obtained by the IceCube~(IC)~\cite{IceCubeCollaborationSS:2025jbi}, Pierre Auger Observatory~(Auger)~\cite{PierreAuger:2023pjg}, and Baikal-GVD~\cite{Baikal-GVD:2025kbe} experiments are well below the value of the neutrino flux measured by KM3NeT.
At the same time, all modern high-energy neutrino experiments can be considered a unified ``global neutrino observatory''. In this case, a joint fit of the flux yields a p-value of $0.5\%$~($2.6\sigma$) for a single detection in KM3NeT and no detections in the IceCube and Auger experiments~\cite{km3net-global-neutrino-landscape}.
The corresponding neutrino flux is significantly lower than that derived from the KM3NeT-only observation.
Furthermore, anomalous UHE events were also detected by the ANITA neutrino experiment~\cite{ANITA:2020gmv}, but their interpretation as UHE neutrinos is questionable~\cite{ANITA:2021xxh}.

The origin of the neutrino KM3-230213A is debatable. 
In the series of papers that followed the detection report, the KM3NeT collaboration discussed its possible galactic~\cite{KM3NeT:2025aps}, blazar~\cite{KM3NeT:2025bxl}, and cosmogenic~\cite{KM3NeT:2025vut} origins.
All of these explanations meet some problems.
The KM3-230213A event was also used for Beyond the Standard Model interpretations~\cite{Dev:2025czz,Brdar:2025azm,Palmisano:2025abd,Jho:2025gaf,Borah:2025igh,Farzan:2025ydi} and constraints~\cite{KM3NeT:2025mfl, Satunin:2025uui, Aloisio:2025nts, He:2025bex}.
In this paper, we discuss the possible cosmogenic origin of the KM3-230213A neutrino. 

Cosmogenic neutrinos are high-energy neutrinos produced by UHECRs during propagation through the interstellar medium. Therefore, to consider the neutrino of this origin, we first need to discuss UHECR.
Despite many years of studies with different experiments, the origin of UHECRs remains an open question.
It is assumed that UHECRs have an extragalactic origin because the galactic magnetic fields are not strong enough to confine them and their observed distribution in the sky is quite isotropic~\cite{AlvesBatista:2019tlv}.
A related problem that has not been solved yet is UHECR mass composition~\cite{AlvesBatista:2019tlv}.
Knowledge of the composition would help in deciphering UHECR origin.
The standard methods of inferring UHECR mass composition are based on the analysis of the development of extensive air showers (EAS) in the atmosphere of Earth. The results of these methods are affected by the stochasticity of air showers and by the uncertainties in their measurements and modeling. 
Thus, the results of the standard UHECR composition inference obtained by two largest modern experiments, Telescope Array~(TA)~\cite{TelescopeArray:2012uws} and Pierre Auger Observatory~(Auger)~\cite{PierreAuger:2015eyc}, are in tension with each other~\cite{PierreAuger:2014gko, TelescopeArray:2018xyi}. 
There are also alternative methods of UHECR mass composition inference, either from surface detector measurements~\cite{TelescopeArray:2018bep, PierreAuger:2017tlx, PierreAuger:2024nzw, PierreAuger:2024flk} or from UHECR anisotropy~\cite{TelescopeArray:2024oux, TelescopeArray:2024buq}, but the former is affected by even larger EAS uncertainties than the standard method, while the latter is affected by uncertainties in UHECR flux modeling, so that its results are robust only at the highest energies $E > 100$~EeV.

The cosmogenic neutrino flux is produced in $p\gamma$ processes and is sensitive to the UHECR mass composition. 
The inference of the UHECR mass composition using the cosmogenic neutrino flux is not affected by the uncertainties accompanying its derivation from EAS measurements.
This advantage motivated the intensive development of cosmogenic neutrino models in recent years~\cite{vanVliet:2019nse, Ehlert:2023btz, Muzio:2023skc}, as well as their experimental searches~\cite{PierreAuger:2019ens, IceCubeCollaborationSS:2025jbi, TelescopeArray:2019mzl}, including proposals for future giant experiments~\cite{GRAND:2018iaj, IceCube-Gen2:2020qha, POEMMA:2020ykm}.
The inverse approach is also viable: assuming some UHECR flux model, one can predict the cosmogenic neutrino flux and estimate its compatibility with the neutrino measurements. This approach was applied to the KM3-230213A neutrino measurement in several studies~\cite{KM3NeT:2025vut, Li:2025tqf, Das:2025vqd, Cermenati:2025ogl}.
It is remarkable that all of them were based on UHECR flux models inspired by the Auger composition measurements. In this study, we consider KM3-230213A neutrino as cosmogenic, assuming minimal models of the UHECR flux dominated by light nuclei, as inferred by the TA experiment~\cite{Bergman:2021djm}.
In these models, we also compute the fluxes of cosmogenic cascade gamma-rays in the MeV–GeV range and cosmogenic UHE gamma-rays and confront the former with the Fermi-LAT IGRB data~\cite{Fermi-LAT:2014ryh} and the latter with Auger and TA UHE diffuse gamma-ray data~\cite{PierreAuger:2024ayl, PierreAuger:2022aty, TelescopeArray:2018rbt, Kharuk:2023gts}.

The paper is organized as follows. In Sec.~2, we briefly introduce the neutrino data of different experiments that we are using; in Sec.~3, we describe the UHECR flux models we use and our setup for simulations of UHECR propagation and cosmogenic flux generation; in Sec.~4, we compute the cosmogenic neutrino flux according to our flux models and compare our predictions with the neutrino data of various experiments and their combinations; in Sec.~5, we compute the cosmogenic gamma-ray flux in the same UHECR models and compare it with the gamma-ray data; in Sec.~6, we compare our results with other studies and discuss the prospects for future tests of the considered UHECR models.

\paragraph{2. Ultra-high energy neutrino data.}\label{sec:nu_data}
In this study, we consider the results of ultra-high energy neutrino search of several experiments, such as KM3NeT~\cite{KM3NeT:2025npi}, IceCube~9 years~\cite{IceCube:2018fhm}, and IceCube 12~years~\cite{IceCubeCollaborationSS:2025jbi}, Pierre Auger~\cite{PierreAuger:2023pjg} and Baikal-GVD~\cite{Baikal-GVD:2025kbe}.
The operation time of these experiments is given in Table~\ref{tab:operation_time}.

\begin{table}
    \centering
    \begin{tabularx}{\linewidth}{YY}
        \hline
        \hline
        {Experiment}      & {Operation time}     \\
        \hline
        KM3NeT            & $335$ days           \\
        Pierre Auger      & $18$ years           \\
        IceCube~$9$~years  & $3142.5$ days        \\
        IceCube~$12$~years & $4605$ days          \\
        Baikal-GVD        & $26.8$ cluster-years \\
        \hline
        \hline
    \end{tabularx}
    \caption{Operation time of neutrino experiments. Note that Baikal-GVD detector consists of several independent clusters. As this telescope was expanding during the operation, the convenient units for operation time of Baikal-GVD are cluster-years.}
    \label{tab:operation_time}
\end{table}

\begin{figure}[!ht]
    \centering
    \input{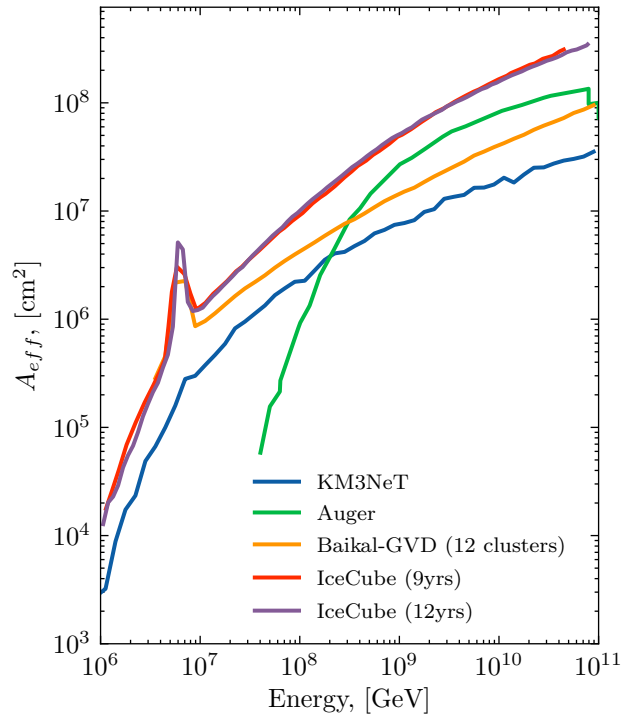}
    \caption{Effective areas of the experiments used in this work. These are full-sky averages, summed over all neutrino flavors, and averaged over both neutrinos and antineutrinos.}\label{fig:effarea}
\end{figure}

To estimate the number of expected events for an experiment, we need to know detector exposure~$\mathcal{E}(E)$, which depends on operation time $T$ and effective area $A_{eff} (E)$:
\begin{equation}
\mathcal{E}(E) = 4\pi \cdot T \cdot A_{eff} (E)
\label{eq:exposure}
\end{equation}
In our case, the effective area $A_{eff} (E)$ is sky-averaged, summed over all neutrino flavors, and averaged over both neutrinos and antineutrinos. The effective areas for the experiments are shown in Fig.~\ref{fig:effarea}. Apart from KM3NeT, these experiments did not detect neutrino events in the energy range discussed in this study ($72\; \mbox{PeV} - 2.6\; \mbox{EeV}$).

\paragraph{3. UHECR flux models and simulation of cosmogenic neutrinos.}\label{sec:models_sim}
Cosmogenic neutrinos are produced during the propagation of UHECRs through the cosmological photon background, i.e. the extragalactic background light (EBL) and the cosmic microwave background (CMB)~\cite{Berezinsky:1970xj,Stecker:1973,Hill:1983xs}:
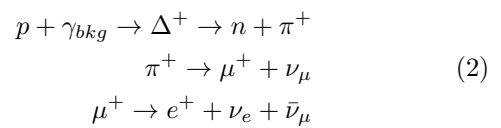
\begin{equation}
\begin{split}
    p+\gamma_{bkg}\rightarrow\Delta^+\rightarrow n+\pi^+\\
    \pi^+\rightarrow\mu^++\nu_\mu\\
    \mu^+\rightarrow e^++\nu_e+\bar{\nu}_\mu
\end{split}
\label{eqn:neutrino-production}
\end{equation}

$\Delta^+$ decay can produce a proton instead of a neutron:
\begin{equation}
    p+\gamma_{bkg}\rightarrow\Delta^+\rightarrow p+\pi^0
    \label{eqn:proton-production}
\end{equation}

This proton still has ultra-high energy. It can, in turn, interact with the CMB and EBL photons (see Eq.~\eqref{eqn:neutrino-production}) and produce more neutrinos.

$\pi^0$ meson from the decay of $\Delta^+$ (Eq.~\eqref{eqn:proton-production}) almost instantly decays into two photons: $\pi^0\rightarrow\gamma\gamma$. These photons are very energetic ($E > 10^{17}$~ eV) and therefore are called ultra-high~energy (UHE) gamma-rays. UHE gamma-rays generated relatively close to Earth ($\leq$ few Mpc~\cite{Bhattacharjee:1999mup}) have a chance to evade interactions and reach the Earth, conserving their high energy. 
Their number should be low, and right now there are only upper limits on their flux~\cite{PierreAuger:2024ayl, PierreAuger:2022aty, TelescopeArray:2018rbt}. UHE gamma rays from more distant sources interact with background photons and produce an electromagnetic cascade: $\gamma+\gamma_{bkg}\rightarrow e^+e^-$. There should also be a contribution to $e^{\pm}$ cascades from UHE protons (Eq.~\eqref{eqn:neutrino-production},~\eqref{eqn:proton-production}) that lose energy by pair production: $p+\gamma\rightarrow p+e^+e^-$.  The energies of photons in the resulting cascades span from~$\sim$~GeV to hundreds of GeV, and they are detectable by gamma-ray observatories~\cite{Berezinsky:2016feh}.

The expected cosmogenic neutrino flux and diffuse UHE photon flux depend strongly on the composition of UHECRs, since protons produce many pions, while heavy nuclei ($A>1$) tend to break up into lighter fragments rather than producing many pions directly. The evolution of the UHECR sources with redshift is also very important, because more numerous or brighter sources at high redshift result in a higher flux at Earth~\cite{kotera-allard}. The latter only applies to cascade photons, since UHE photons are produced relatively close to Earth i.e. at low redshifts. 
Lastly, the neutrino flux strongly depends on the maximum acceleration energy at sources, because it translates to the number of protons with energies high enough for photopion production (see Eq.~\eqref{eqn:neutrino-production},~\eqref{eqn:proton-production})~\cite{minimal-cosmogenic-neutrinos}.

\textbf{UHECR flux models.}
This research is based on the UHECR source models presented in the study~\cite{Bergman:2021djm}. 
These models were derived from the combined fit of the TA surface detector (SD) spectrum measurements~\cite{Tsunesada:2017aaq} and the TA fluorescence detector (FD) stereo composition measurements~\cite{Bergman:2017ikd}.
In this section, we briefly discuss the main properties and methodology of the TA UHECR models.

It is assumed that the UHECR sources are identical and uniformly distributed, following the sources evolution function~(SE)~\cite{vanVliet:2019nse}:
\begin{equation}
\label{eqn:evolution}
    \text{SE}(z) = \begin{cases}  {(1+z)}^3 & \text{if } z < 1.5 \\  {2.5}^3 & \text{else }\end{cases}
\end{equation}
where the redshift $z$ extends up to $z_{max} = 4$.

The sources emit a mixture of five nuclei~$A$: protons~(\textit{p}), helium~($^4$\textit{He}), nitrogen~(${}^{14}$\textit{N}), silicon~(${}^{28}$\textit{Si}), and iron~(${}^{56}$\textit{Fe}) with given fractions $f_A$
and spectra with identical power-law slopes~$\gamma$ and a rigidity-dependent broken exponential cutoff. 
Thus, the differential energy spectrum of the emitted nuclei $A$~${dN_A}/{dE}$ is the following:
\begin{gather*}
    \frac{dN_A}{dE} = f(E) \cdot     \begin{cases}
        1 & \text{if } E < Z_AR_{max} \\  
        \exp{\left(1 - \frac{E}{Z_AR_{max}}\right)} & \text{else }
    \end{cases} \\
    f(E) = J_0 f _A \cdot {\left(\frac{E}{\text{1 EeV}}\right)}^{-\gamma}
\end{gather*}
where $J_0$ is the spectrum normalization constant, $Z_A$ is a charge of $A$ nuclei, and $R_{max}$ is a maximum rigidity.
Therefore, the parameters of the UHECR flux are the following: $\gamma$, $R_{max}$, and five nuclei fractions $f_A$ that sum up to unity. The general setup of this model follows the similar study of the Pierre Auger collaboration~\cite{PierreAuger:2016use}.

The model is built using only the TA spectrum and $X_{\rm max}$ data for $E > 10^{18.7}$~EeV, i.e., above the ankle. This implies that the UHECR flux at lower energies originates from the separate source population and is not described by the considered model. 
Although a low-energy UHECR component should exist, it is not necessarily extra-galactic and not necessarily light. Both of these options would lead to negligible cosmogenic neutrino flux. Therefore, in this study we do not use any model for the sub-ankle component and focus on the high-energy component as a minimal model to describe the measurements of high-energy neutrino experiments.
Also, we would like to stress that the evolution of the sources in the considered flux model is fixed.

\begin{table}
    \centering
    \begin{tabularx}{\linewidth}{lYYc}
        \hline
        \hline
        {Model} & $\gamma$ & $R_{max}$ & $f_A$ \\
        \hline
        Best-fit & $2.06$ & $182\,\text{EV}$ & \textit{He}: 0.992, \textit{Fe}: 0.008\\
        Local min. & $0.78$  & $15.8\,\text{EV}$ & \textit{p}: 0.971, \textit{Fe}: 0.029 \\
        \hline
        \hline
    \end{tabularx}
    \caption{The parameters of the used UHECR source models.
    Fractions of the nuclei species that are not shown in the table are equal to zero.}
    \label{tab:model_params}
\end{table}

\textbf{Simulation setup.}
Technically, nuclei propagation is simulated using the CRPropa~(version~$3.2.1$) framework~\cite{AlvesBatista:2022vem}.
The simulation takes into account the interaction of UHECRs with the CMB
and uses the Gilmore model~\cite{Gilmore:2011ks} of the infrared background (IRB) to calculate photo-pion production, photo-nuclear disintegration, and electron-positron pair production. It also calculates electromagnetic interactions such as electron pair production and inverse Compton scattering.

\begin{figure}[!t]
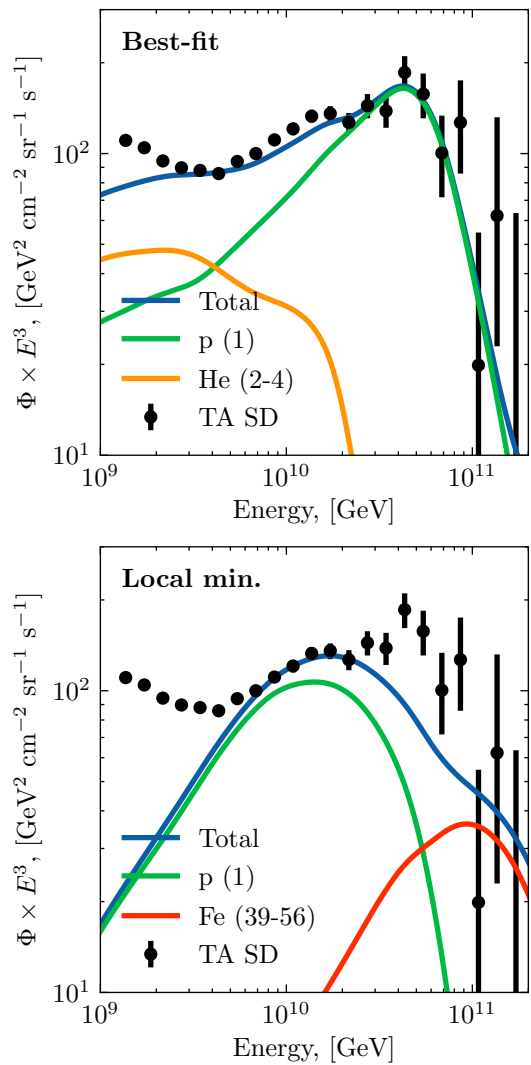

    \centering
    \begin{subfigure}[b]{0.5\textwidth}
        \centering
        \input{assets/fig_ta_best.pgf}
    \end{subfigure}
    ~
    \begin{subfigure}[b]{0.5\textwidth}
        \centering
        \input{assets/fig_ta_local.pgf}
    \end{subfigure}
    \caption{Simulated spectra of UHECR mass components according to the TA flux models (solid lines).
    The upper and lower plots correspond to the best-fit and local min. models, respectively. 
    Black dots with error bars --- 
    all-particle UHECR spectrum measured with the TA~SD.}
    \label{fig:bergman}
\end{figure}

In our study, we consider two results (i.e., two models) for a combined fit to the TA cosmic-ray spectrum and composition.
One of them corresponds to the best minimum, and another --- to the local one~\cite{Bergman:2021djm}.
We call them ``best-fit'' and ``local min.'', respectively, for simplicity.
Their parameters are described in Table~\ref{tab:model_params}.
We have simulated CR propagation for one million events in both models.
In Fig.~\ref{fig:bergman}, the measured TA SD spectrum~\cite{Tsunesada:2017aaq} is compared with those computed for two models.
In this figure, the measured spectrum is shifted by $\Delta \log_{10}{E} = -0.12$ as prescribed by the fit of Ref.~\cite{Bergman:2021djm}, to take into account possible systematic uncertainties in the measured energy.
The total hadronic fluxes computed with CRPropa are normalized to the TA spectrum at an energy point of $20$\,EeV.
We checked that the UHECR components spectra from the original study~\cite{Bergman:2021djm} are well reproduced in our simulations. 

One can see from Fig.~\ref{fig:bergman} that the local min. model does not reproduce the measured spectrum at the highest energies.
The goodness-of-fit of the local min. model is also much worse than that of the best-fit model. However, we think it is interesting to consider this model on equal grounds with the best-fit model. The part of the spectrum with higher energies, which is not reproduced in the local min. model, can be attributed to a distinct close UHECR source. Indeed, the recent TA study shows the significant declination dependence of the all-particle spectrum:  there is a ``bump'' at the highest energies in the Northern part of the sky~\cite{TelescopeArray:2024tbi}. This observation can be interpreted as a contribution of the local source~\cite{Plotko:2022urd}. 
There are independent arguments in favor of this interpretation. 
First, there is the evidence for the close UHECR source~\cite{Kuznetsov:2023jfw} that follows from the recent observation of the extreme energy particle by the TA~\cite{TelescopeArray:2023sbd}. Second,
the heavy composition at the highest energies in the local min. model
is compatible with recent TA results on the heavy composition in this energy range~\cite{TelescopeArray:2024oux, TelescopeArray:2024buq}.
Therefore, we consider the local min. model of special interest for research.

\paragraph{4. Predictions of neutrino flux and comparison with the data.}
\label{sec:preds_vs_data}
We simulate the propagation of nuclei emitted by UHECR sources for two models described in the previous Section, using the Monte Carlo framework CRPropa.
These simulations also incorporate production of cosmogenic neutrinos through the interaction of cosmic rays with EBL and CMB.
The output includes the observed cosmogenic neutrino flux (electron and muon components), the UHECR flux, and also gamma-ray flux.
All the output fluxes are scaled by the same factor, which is derived from the fit of the all-particle UHECR flux to the TA observed spectrum, as described in the previous Section.
To take the oscillations of neutrino flavors into account, we
additionally average over all neutrino flavors and sum over both neutrinos and antineutrinos to get the resulting cosmogenic neutrino flux: $\Phi^{1f}_{\nu + \bar{\nu}}(E)$.
It is important to note that UHECR sources with $z>1$ make a small contribution to the cosmic ray spectrum, but their contribution to the neutrino flux is important.

\begin{figure}[!t]
    \centering
    \input{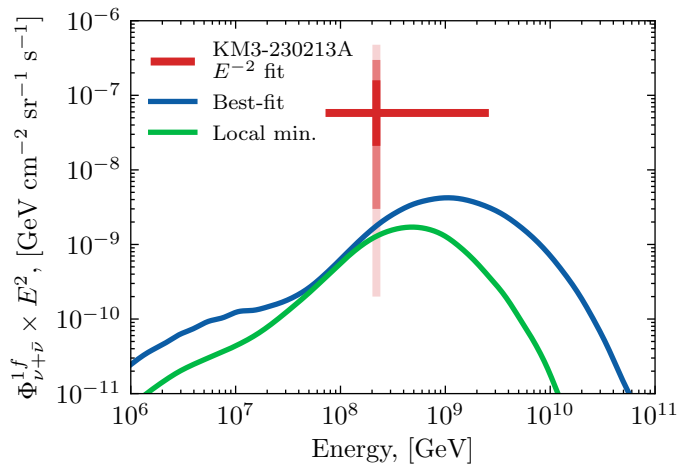}
    \caption{
        Energy-squared per-flavour neutrino fluxes, assuming neutrino equipartition ($\nu_e:\nu_\mu:\nu_\tau = 1:1:1$).
        Red and olive crosses denote KM3NeT-only measurement~\cite{KM3NeT:2025npi} and combined $E^{-2}$ fit~\cite{Baikal-GVD:2025kbe}.
        Vertical bars represent the $1\sigma$, $2\sigma$, $3\sigma$ and $1\sigma$ Feldman–Cousins confidence intervals~\cite{Feldman:1997qc} of these estimates respectively. 
        Blue and green solid lines show the fluxes for the best-fit and local min. models, respectively.
    }
    \label{fig:neutrino}
\end{figure}

A first simple test of our flux models involves a comparison with the so-called KM3NeT-only flux~\cite{KM3NeT:2025npi} and the combined $E^{-2}$ fit flux~\cite{Baikal-GVD:2025kbe}.
KM3Net-only flux refers to the flux $\Phi(E)\propto E^{-2}$ needed to obtain exactly one neutrino event within the central 90\% energy range associated with the KM3-230213A event.
Built by analogy, the combined fit additionally implies the absence of observations in the same energy range from the Auger, IceCube~12 years, and Baikal-GVD experiments.
Coarse visual inspection of Fig.~\ref{fig:neutrino} shows that our fluxes reside within the $3\sigma$ interval of the KM3NeT-only $E^{-2}$ flux and of the combined fit flux, so the tension is not strong.

For a more thorough comparison with neutrino experiments, we need to estimate the number of high-energy neutrino events expected in a given model and compare it with the number of observed events. 
To calculate the number of events expected for a given neutrino experiment, $n_{pred}$, we need to integrate the experiment exposure~$\mathcal{E}(E)$~\eqref{eq:exposure} with the cosmogenic neutrino flux $\Phi^{1f}_{\nu + \bar{\nu}}(E)$ predicted by a given model:
\begin{equation}
    n_{pred} = \int_{E_{min}}^{E_{max}} \mathcal{E}(E) \cdot \Phi^{1f}_{\nu+\bar{\nu}}(E) dE \,.
\label{eq:exp_num}
\end{equation}
\noindent
Here we use the 90\% confidence level interval for KM3-230213A energy (72~PeV, 2.6~EeV) as integration limits $E_{min}$, $E_{max}$.

The estimated numbers of events are shown in Table~\ref{tab:neutrino_results_best}.
The local min. model gives approximately two times lower estimate for the number of events than the best-fit model. Both models predict at least one event for IceCube~9~years and IceCube~12~years, a slightly lower number of events for Auger, and significantly less than one event for KM3NeT and Baikal-GVD.
Both models are compatible with the KM3NeT-only measurement within about $2\sigma$.
At the same time, we can estimate the consistency of our flux models with the results of ``global neutrino observatory'' using combined $p$-value for all experiments.
The combined results are also compatible with both best-fit and local min. models at about $2\sigma$ level.
The details of the computations of the combined significances are given in the Appendix.

\begin{table*}[]
    \centering
    \begin{minipage}[t]{85mm}
    \begin{tabularx}{0.94\columnwidth}{cYYYc}
        \multicolumn{5}{c}{Best-fit model} \\
        \hline
        \hline
         Experiment & $n_{pred}$ & $n_{obs}$ & $p$-value & Significance \\
         \hline
         KM3NeT & $0.037$ & 1 & $0.036$ & $1.80\sigma$ \\
         Auger & $1.67$ & 0 & $0.188$ & $0.89\sigma$ \\
         IC 9 years & $1.90$ & 0 & $0.150$ & $1.04\sigma$ \\
         Baikal-GVD & $0.16$ & 0 & $0.849$ & $-$ \\
         IC 12 years & $2.97$ & 0 & $0.051$ & $1.63\sigma$ \\
        \hline
         Combined (MC) & $-$ & $-$ & $0.019$ & $2.09\sigma$ \\
         Combined ($\chi^2$) & $-$ & $-$ & $0.0064$ & $2.49\sigma$ \\
        \hline
        \hline
    \end{tabularx}
    \end{minipage}
    \begin{minipage}[t]{85mm}
    \begin{tabularx}{0.94\columnwidth}{cYYYc}
        \multicolumn{5}{c}{Local min. model} \\
        \hline
        \hline
         Experiment & $n_{pred}$ & $n_{obs}$ & $p$-value & Significance \\
         \hline
         KM3NeT & $0.019$ & 1 & $0.019$ & $2.08\sigma$ \\
         Auger & $0.69$ & 0 & $0.504$ & $-$ \\
         IC 9 years & $0.90$ & 0 &  $0.405$ & $0.24\sigma$ \\
         Baikal-GVD & $0.08$ & 0 & $0.920$ & $-$ \\
         IC 12 years & $1.43$ & 0 & $0.240$ & $0.71\sigma$ \\
        \hline
         Combined (MC) & $-$ & $-$ & $0.035$ & $1.81\sigma$ \\
         Combined ($\chi^2$) & $-$ & $-$ & $0.035$ & $1.81\sigma$ \\
        \hline
        \hline
    \end{tabularx}
    \end{minipage}
    \caption{
     Predictions of two UHECR flux models and a comparison with the results of different neutrino experiments.
     $n_{pred}$ --- number of expected neutrino events in the KM3-230213A energy range (90\% confidence interval), $n_{obs}$ --- number of observed neutrino events. $p$-values for separate experiments are Poissonian, and statistical significances are one-sided. Combined $p$-values and statistical significances are calculated for the KM3NeT, Auger, Baikal-GVD, and IC 12 years experiments (see Appendix).
    }
    \label{tab:neutrino_results_best}
\end{table*}

\paragraph{5. Predictions of gamma-ray flux and comparison with the data.}\label{sec:gammas}
~

\begin{figure}[!ht]
    \centering
    \input{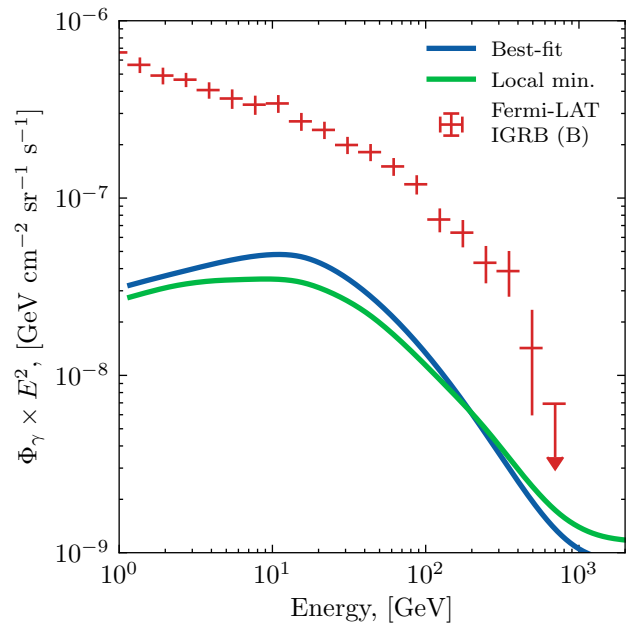}
    \caption{Energy spectrum of cascade photons. 
    Blue and green solid lines represent the estimated flux for the best-fit and local min. models, respectively. 
    Red points with error bars represent the Fermi IGRB measurements for galactic foreground model B~\cite{Fermi-LAT:2014ryh}.}\label{fig:fermi}
\end{figure}

\textbf{Cosmogenic cascade gamma-rays and Fermi-LAT IGRB.}\label{sec:gammas_fermi}
Another important constraint on UHECR source models comes from measurements of the isotropic diffuse gamma-ray background~(IGRB)~\cite{Berezinsky:2016jys}.
In particular, the results of the Fermi-LAT observatory, in the energy range from 100~MeV to 820~GeV are suitable for this purpose~\cite{Fermi-LAT:2014ryh}.
These constraints are especially useful for proton-dominated UHECR models, as protons produce cascades more efficiently than nuclei.
Proton energy is converting to electromagnetic cascades through photo-pion and electron-positron production on the CMB and the EBL~\cite{Berezinsky:2016feh}. Fermi-LAT can detect gamma rays from this $e^\pm$-cascade.

In Fig.~\ref{fig:fermi} we show the flux of cascade gamma-rays produced by the best-fit and local min. models. 
The resulting fluxes for both models lie well below the Fermi-LAT IGRB estimation.
It is worth noting that the expected contribution to the IGRB could be higher once the low-energy UHECR component is assumed (see e.g. Ref.~\cite{Cermenati:2025ogl}). 
We do not consider this possibility in the present study. 
Therefore, we can only conclude that our models for the high-energy UHECR component do not contradict the Fermi-LAT IGRB measurements.

\textbf{Cosmogenic UHE gamma-rays.}
Ultra‑high energy~(UHE) diffuse gamma-rays with $E > 10^{17}$~eV are produced by interactions of UHECRs with background radiation via the Greisen–Zatsepin–Kuzmin (GZK) mechanism~\cite{Wdowczyk:1971jk, Gelmini:2005wu}. 
These photons subsequently can initiate electromagnetic cascades and lose energy during propagation. The expected flux of UHE gamma-rays is small, and no detections have been reported to date. 
Nevertheless, experimental upper limits from TA~\cite{TelescopeArray:2018rbt, Kharuk:2023gts} and Auger~\cite{PierreAuger:2022aty, PierreAuger:2024ayl}
are starting to constrain the most optimistic (i.e. protonic) UHECR flux models.

\begin{figure}[!ht]
    \centering
    \input{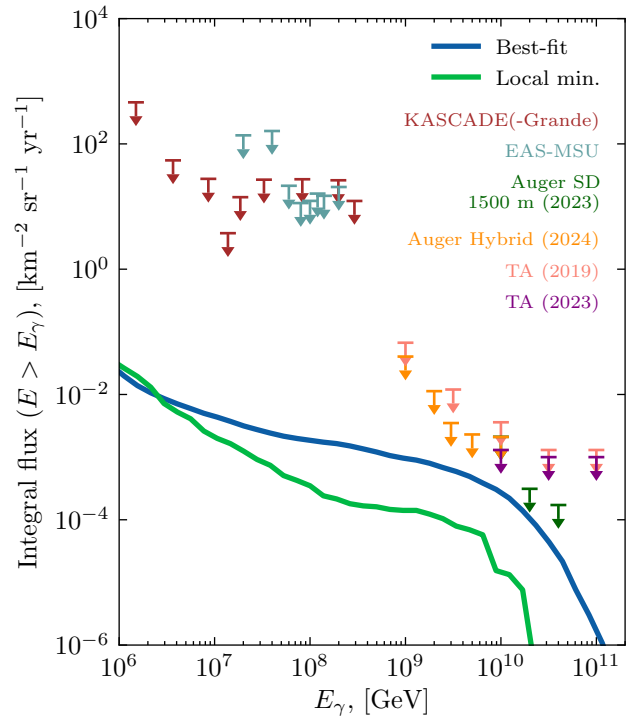}
    \caption{Upper limits on the integral UHE gamma ray flux. Here we compare predictions of the models (solid lines) with the observations (markers with downward arrows). Upper limits from Auger~\cite{PierreAuger:2024ayl, PierreAuger:2022aty} are marked in green and orange, limits from TA~\cite{TelescopeArray:2018rbt, Kharuk:2023gts} are marked in pink and purple, limits from KASCADE(-Grande)~\cite{KASCADEGrande:2017vwf} are marked in red and limits from EAS-MSU~\cite{Fomin:2017ypo} are marked in teal. Blue and green solid lines correspond to the best-fit and local min. models, accordingly.
    }
    \label{fig:integral_flux}
\end{figure}

Here we test the TA UHECR flux models for consistency with these experimental results. Fig.~\ref{fig:integral_flux} shows the integral gamma-ray flux predictions for the best-fit and local min. models together with the observational upper limits. 
One can see that the model predictions are below the limits, meaning that neither model is excluded by the present UHE gamma-ray constraints. 
At the same time, the limits at the highest energies are close to the flux predicted in the best-fit model, suggesting that it can be probed by UHECR experiments in the near future.
We also note that, because of the relatively small attenuation length of the UHE gamma-rays, their measurement can test only the local behavior of the UHECR flux, while the source evolution is unconstrained by these limits. However, for the flux models discussed in this study, this is not important as we are limiting ourselves to the analysis of the basic version of the models with conservative source evolution.

It is important to note that our simulations only include interactions with the CMB and EBL.
However, the predicted photon flux can also be affected by the universal radio background~(URB)~\cite{Protheroe:1996si}, that is not included in our simulations.
Therefore, these predictions need to be interpreted as an optimistic expectation of the magnitude of the UHE gamma-ray flux.
It was shown in Ref.~\cite{Gelmini:2022evy} that taking URB into account decreases the UHE gamma-ray flux (e.g. by $\sim 3$ times at $E \approx 10^{10}$~GeV).

\paragraph{6. Discussion.}\label{sec:discussion}
In this study, we have shown that the models of UHECR flux that describe the energy spectrum and $X_{\rm max}$ data measured by the Telescope Array experiment naturally predict the flux of cosmogenic neutrino that is reasonably compatible with the KM3NeT measurement of the KM3-230213A ultra-high energy neutrino event.
Namely, the models predict neutrino fluxes that are smaller than the KM3NeT-only measurement, but compatible with it at $\sim 2 \sigma$ level. At the same time, these models yield the expectation of $1-2$ neutrino events in the same energy range for other high-energy neutrino experiments: IceCube, Auger, and Baikal-GVD, while no events were observed by them. The combined ``measurement'' of all these experiments (including KM3NeT) is compatible with our expectations at the same $\sim 2 \sigma$ level.
The UHECR composition in both models is dominated by protons and light nuclei. The source evolution with redshift in these models is standard, with $m=3$ at $z < 1.5$ and no sources at $z>4$.

There were several studies on the possible cosmogenic origin of the KM3-230213A event~\cite{KM3NeT:2025vut, Muzio:2025gbr, Das:2025vqd, Cermenati:2025ogl}. The study by the KM3NeT collaboration itself~\cite{KM3NeT:2025vut} uses the UHECR model with intermediate nuclei composition based on the results of the Pierre Auger experiment~\cite{PierreAuger:2016use}.
This leads to the necessity of introducing extreme source evolution (most of the neutrino flux comes from very high redshifts) to reconcile the model with the observation of the KM3-230213A event.
Another possibility of enhancing the cosmogenic neutrino flux to the level of the KM3-230213A measurement, while considering the intermediate nuclei composition models, is to add a separate proton component to the UHECR flux~\cite{Muzio:2023skc, Ehlert:2023btz, Muzio:2025gbr, Cermenati:2025ogl}. This component should have a different injection spectrum and should originate from another source population.
However, as it was shown in Ref.~\cite{Cermenati:2025ogl}, the addition of this component leads to an overproduction of IGRB at $10-100$~GeV, making this scenario implausible.

In this context, it is remarkable that the minimal models we consider demonstrate a reasonable compatibility with the neutrino data from all the high-energy neutrino experiments without any tuning. That is, no extreme evolution of sources, no contribution from high redshift sources, and no additional proton source population are needed.

We also tested the predictions of cosmogenic gamma-ray fluxes from the same UHECR flux models against the gamma-ray constraints from Fermi-LAT, Auger, and TA. The expected fluxes do not exceed the existing gamma-ray limits. 
The predictions for UHE gamma rays could be probed by the Auger and TA experiments if the efficiency of their gamma-ray search analyses were improved. For example, the sensitivity of the Auger analysis should be increased by $\sim 3$ times to start probing the best-fit model at $2 \cdot 10^{10}$~GeV, while the sensitivity of the TA analysis should be increased by $\sim 5$ times at $10^{10}$~GeV. At the same time, the predictions of the local min. model are unreachable by the recent UHECR experiments. Still, there is hope for future cosmic-based UHECR experiments for this purpose.

The code for this paper is available at GitHub\footnote{\href{https://github.com/82492749123082/KM3-230213A_UHECR_TA}{https://github.com/82492749123082/KM3-230213A\_UHECR\_TA}}.

We would like to thank D.~Bergman for the clarifications about the Telescope Array UHECR flux models
and also S.~Troitsky, O.~Kalashev, G.~Rubtsov, and P.~Satunin for the helpful discussions.

\paragraph{7. Funding.}
This work was supported by the Russian Science Foundation grant \mbox 25-12-00111.

\paragraph{8. Conflict of interest.}
The authors of this work declare that they have no conflicts of interest.

\paragraph{Appendix: calculation of the combined significance.}\label{sec:combined_significance}
To evaluate the compatibility of a given model with neutrino experiments, we employ the following procedure.
We compare the observed number of neutrino events ($n^{obs}_i$) from each experiment with the number predicted by the model ($n^{pred}_i$)  for the same experiment 
in the central 90\% energy range of the KM3-230213A event ($72 \text{\,PeV} - 2.6 \text{\,EeV}$). 
The likelihood ($\mathcal{L}$) is then calculated as the product of the individual Poisson probabilities for each experiment:
\begin{equation}
\mathcal{L} = \prod_{i} \mathbb{P}(n^{obs}_i, n^{pred}_i),
\end{equation}
Here, the index $i$ corresponds to a specific experiment (KM3NeT, Auger, Baikal-GVD, and IC 12 years), 
and $\mathbb{P}(n^{obs},n^{pred})$ is the Poisson probability of observing $n^{obs}$ events when $n^{pred}$ events are predicted by the model.

To quantify the model's compatibility with the data, we compute a combined $p$-value, 
defined as the probability of obtaining a likelihood equal to or lower than the one observed. 
This $p$-value is then converted into a one-sided significance.

The $p$-value is determined using two complementary methods: a Toy Monte Carlo (MC) simulation and an approximation based on Wilks'~\cite{Wilks:1938dza} theorem.
For the first method, we generate a dataset of one million samples for each experiment, 
where the event counts are drawn from a Poisson distribution with the mean value predicted by the model. 
The $p$-value is then computed from this distribution of simulated likelihoods.
Alternatively, we use the following relation, 
which states that the double difference in the logarithm of the likelihoods follows a chi-squared ($\chi^2$) distribution:
\begin{equation}
\chi^2 \sim -2 \left(\ln \mathcal{L} - \ln \mathcal{L}_0 \right),
\end{equation} 
where $\mathcal{L}_0$ is the maximum possible likelihood, corresponding to a saturated model where the predicted number of events perfectly matches the observed number for each experiment ($n^{pred}_i = n^{obs}_i$). In our case, the $\chi^2$ distribution has 4 degrees of freedom (d.o.f.), corresponding to the four independent experiments included in the test. 
The similar approach was used in the KM3NeT study~\cite{km3net-global-neutrino-landscape} to quantify the compatibility of the combined measurements of several experiments with the neutrino flux models. Therefore, our results in terms of ($\chi^2$): combined p-values and corresponding statistical significances can be directly compared with those of the mentioned KM3NeT study. However, we need to note that this approach is an approximation; the results of the MC method are more precise.

It is important to note that this combined test uses independent experimental results. 
For this reason, we include the IceCube 12-year results and exclude the earlier IceCube 9-year results, as they are not statistically independent.

\bibliographystyle{apsrev4-2}
\bibliography{biblio.bib}

\end{document}